\documentclass[a4paper]{jpconf}
\usepackage[hyperindex,
            colorlinks=true,linktocpage=false,
            bookmarks=false,
            breaklinks=true,
            pdftitle={Random template placement and prior information},
            pdfsubject={},
            pdfauthor={Christian Roever},
            pdfkeywords={},
            linkcolor=black,citecolor=black,urlcolor=black]{hyperref}

\usepackage{dsfont}       
\usepackage{graphicx}

\providecommand{\solarmass}{\mathrm{M}_{\scriptscriptstyle\odot}}
\providecommand{\likeli}{\mathcal{L}}
\providecommand{\pstar}{p^{\star}}
\providecommand{\rhohat}{\hat{\rho}}
\providecommand{\prob}{\mathrm{P}}
\providecommand{\realline}{\mathds{R}}
\providecommand{\expect}{\mathrm{E}}
\providecommand{\differential}{\mathrm{d}}
\providecommand{\normaldistn}{\mathrm{N}}
\providecommand{\pstarspace}{\mathcal{P}^\star}

\begin{document}

\marginpar{\hspace{-2.45cm}\texttt{AEI-2009-115}}

\title{Random template placement and prior information}

\author{Christian R\"{o}ver}

\address{Max-Plack-Institut f\"{u}r Gravitationsphysik 
         (Albert-Einstein-Institut), 
         Callinstra{\ss}e~38, 30167~Hannover, Germany.}

\begin{abstract}
  In signal detection problems, one is usually faced with the task of
  searching a parameter space for peaks in the likelihood function
  which indicate the presence of a signal.  Random searches have
  proven to be very efficient as well as easy to implement, compared
  e.g. to searches along regular grids in parameter space.  Knowledge
  of the parameterised shape of the signal searched for adds structure
  to the parameter space, i.e., there are usually regions requiring to
  be densely searched while in other regions a coarser search is
  sufficient.  On the other hand, prior information identifies the
  regions in which a search will actually be promising or may likely
  be in vain.  Defining specific figures of merit allows one to
  combine both template metric and prior distribution and
  devise optimal sampling schemes over the parameter space.
  We show an example related to the gravitational wave signal from a
  binary inspiral event.  Here the template metric and prior
  information are particularly contradictory, since signals from
  low-mass systems tolerate the least mismatch in parameter space
  while high-mass systems are far more likely, as they imply a greater
  signal-to-noise ratio (SNR) and hence are detectable to greater
  distances.  The derived sampling strategy is implemented in a Markov
  chain Monte Carlo (MCMC) algorithm where it improves convergence.
\end{abstract}

\section{Introduction}
Signal detection, in gravitational wave detection in particular,
frequently entails the problem of performing a computationally
expensive numerical search over a large parameter space.  The
\textsl{search} here means a search for a peak in the likelihood
function, or another detection statistic, based on the data at hand
and varying the unknown signal parameters.  A peak or a threshold
excess then indicates the presence of a signal
\cite{McDonoughWhalen,WainsteinZubakov}.  Such ``brute-force''
searches may be implemented as \textsl{grid searches}, evaluating the
detection statistic at regularly placed points in parameter space.
Computing the detection statistic usually means evaluating the
\textsl{match} between a \textsl{signal template} and the data; the
spacing between evaluated points in parameter space is then usually
based on a \textsl{template metric} which ensures that all possible
signals (corresponding to points in parameter space) have at least a
certain minimal match with one of the evaluated templates
(corresponding to the grid points).  Instead of using regularly spaced
template banks
(e.g. \cite{CroceEtAl2002,BabakEtAl2006,Prix2006,Cokelaer2007}), the
use of \textsl{random template banks} has recently gained popularity,
as these are often very easily implemented, and have also be shown to
be very efficient, especially in higher dimensions
\cite{Babak2008,MessengerPrixPapa2009,HarryAllenSathyaprakash2009,MancaVallisneri2009}.
Here the idea is to populate the parameter space randomly, but
uniformly with respect to the template metric.

These template placement strategies have by now usually been based on
``\textsl{minimax}'' reasoning, by aiming at minimizing the maximal
(worst-case) mismatch across the whole parameter space.  Once one
takes prior information on the unknown parameters into consideration,
by accounting for a~priori probabilities attached to different regions
of parameter space, a decision-theoretic approach allows us to devise
other strategies, effectively concentrating efforts on the more
promising regions of parameter space in pursuit of a certain
optimality criterion \cite{Berger,Ferguson}.  In fact, a minimax
strategy may often only exist once one imposes hard bounds on the
parameter space (and by that ensuring the existence of an absolute
\textsl{worst case}).

Markov chain Monte Carlo (MCMC) methods are meanwhile widely used for
(Bayesian) parameter estimation in the signal processing stage for
gravitational-wave signals
\cite{ChristensenMeyer1998,UmstaetterEtAl2004,ArnaudEtAl2007a}.  
MCMC algorithms are, first of all, methods for \textsl{stochastic
integration} \cite{MetropolisUlam1949,MCMCinPractice}, although by the
way they work they often behave similarly to \textsl{stochastic
search} algorithms as well.  This is in fact a most welcome property,
as part of the parameter estimation problem is usually also a
search/optimization problem, as, besides integration over the
parameters' posterior distribution, it requires finding the global
mode or secondary modes.  \textsl{Parallel tempering}
\cite{HukushimaNemoto1996,Hansmann1997} is a variety of the
Metropolis-Hastings MCMC algorithm (and a special case of
Metropolis-coupled MCMC algorithm \cite{Geyer1991,MCMCinPractice})
aimed at enhancing these stochastic search capabilities.  This is done
by basically running several MCMC chains in parallel, where
\textsl{tempering} at increasing temperature values is applied to
subsequent chains (as in simulated annealing methods
\cite{KirkpatrickGelattVecchi1983,NumRecipes}), and additional steps
are introduced to allow for communication between chains
\cite{Roever2007-thesis}.  Parallel tempering methods have been
applied to gravitational-wave data analysis for binary inspiral
signals in the context of ground-based
\cite{RoeverMeyerChristensen2007a} and space-based (LISA) measurements
\cite{Roever2007-thesis}, where they have proven advantageous
especially in cases of high SNR and of posterior distributions
exhibiting multiple modes or degeneracies
\cite{VanDerSluysEtAl2008b,RaymondEtAl2009}; they have also been
adopted for the analysis of burst signals \cite{KeyCornish2009} and
utilized for Bayesian evidence computation
\cite{LittenbergCornish2009}.

Among the parallel Markov chains being run at different `temperatures'
within the parallel tempering implementation, the `cool' ones with no
tempering applied produce samples from the posterior distribution for
the stochastic integration part, while the high-temperature chains are
producing samples for the stochastic search.  The question now is how
to set up the algorithm so that the search is most efficient, given
our knowledge of prior and template metric, i.e., our knowledge of
``where the true parameters are (un-) likely to be'', and ``how hard
one needs to look'' across the parameter space.  The problem is of
special interest in the context of binary inspiral signals, as prior
and template metric are particularly contradictory: a~priori one is
most likely do detect an inspiral involving high masses, as these
result in a high-SNR signal that is detectable to a greater distance.
On the other hand, considering the template metric only, one might
want to mostly try low-mass templates, since at low masses the
template's and true signal parameters need to be in very close
agreement in order for them to match, while at high masses greater
discrepancies still yield a good match.  What needs to be defined is
the distribution to sample from in order to find the mode(s) fastest,
which is very similar to setting up a random template bank, the
difference being that one does not settle on some fixed number of
templates, as the MCMC sampler in principle is thought to sample
indefinitely.

In the following Sec.~\ref{sec:BinaryInspiral}, we will introduce the
problem for the case of binary inspiral signals, and
Sec.~\ref{sec:ParallelTempering} briefly introduces the parallel
tempering context.  In Sec.~\ref{sec:DecisionTheory} the general
problem is formulated in decision-theoretic terms and solved for a
particular optimality criterion.  Sec.~\ref{sec:examples} shows some
illustrative examples, and Sec.~\ref{sec:ConlcusionsOutlook}
eventually closes with conclusions and perspectives.

\section{\label{sec:BinaryInspiral}Binary inspiral parameters}
\begin{figure}
\begin{center}
\includegraphics[width=0.98\textwidth]{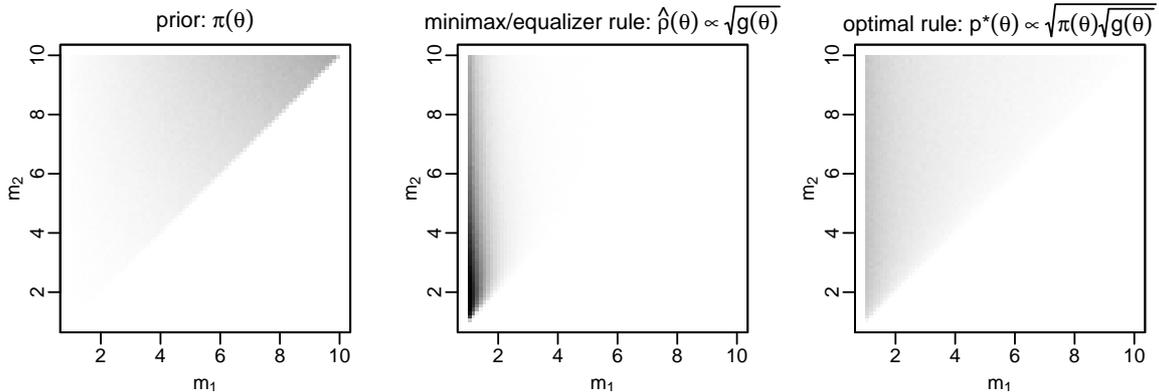}
\end{center}
\caption{\label{fig:inspiralParam}(Marginal) densities of the
distributions $\pi$, $\rhohat$ and $\pstar$ for the two mass
parameters ($m_1$, $m_2$) of a binary inspiral signal.  The prior
(left plot) indicates that high masses are most likely, which is
because they result in stronger signals that are detectable to greater
distances.  The template metric on the other hand implies that low
masses require a dense template spacing (middle plot).}
\end{figure}

In the simplest description, a binary inspiral signal as measured by
ground-based interferometers is determined by 9 parameters: sky
location (declination $\delta$, right ascension $\alpha$), polaristion
($\psi$), companion masses ($m_1$, $m_2$), luminosity distance
($d_L$), time of arrival ($t_c$), phase ($\phi$) and inclination angle
($\iota$).  Assuming some prior distribution for the masses (in the
following simply defined to be uniform, $m_1,m_2\in [1\,\solarmass,
10\,\solarmass]$), and an isotropic distribution of events across
space while folding in the detectability as a function of
signal-to-noise ratio (SNR), one can derive a joint prior distribution
whose marginal distribution of masses is shown in
Fig.~\ref{fig:inspiralParam} \cite{RoeverEtAl2007b,Roever2007-thesis}.
A template metric may be defined following
\cite{OwenSathyaprakash1999,ChronopoulosApostolatos2001}, assuming the
metric to be constant in the space of the \textsl{Newtonian} and
\textsl{1.5~PN chirp times} $\lambda_1$ and $\lambda_2$, which are
functions of the mass parameters.  For the remaining parameters, for
now, we again assume the metric to be uniform ($t_c$, $\log(d_L)$) and
isotropic ($\delta$, $\alpha$, $\psi$, $\iota$, $\phi$).  The implied
distribution in terms of ($m_1, m_2$) following from a uniform spacing
in ($\lambda_1, \lambda_2$) may be derived using the reparametrisation
explicated in \cite{UmstaetterTinto2008}.  This distribution is shown
in Fig.~\ref{fig:inspiralParam}.

\section{\label{sec:ParallelTempering}Parallel tempering}
In the context of Monte Carlo integration, tempering is utilised to
prevent the integration algorithm from getting stuck in local modes of
the distribution from which it is sampling.  A temperature parameter
$T\geq 1$ is introduced, and instead of sampling from the distribution
of actual interest, with density function $f(\theta)$, the modified
distribution
\begin{equation}\label{eqn:plainTempering}
  f_{(T)}(\theta) \;\propto\; f(\theta)^\frac{1}{T}
\end{equation}
is used.  The introduced exponent is supposed to make the distribution
more tractable, as it has a ``flattening'' effect on the density; the
same effect is also taken advantage of in simulated annealing methods
\cite{NumRecipes}.  In the limit of $T\rightarrow\infty$, the density
$f_{(T)}(\theta)$ then approaches a uniform distribution
\cite{MCMCinPractice}.  In the context of posterior inference, when
the target distribution~$f(\theta)$ is the product of
prior~$\pi(\theta)$ and likelihood~$\likeli(\theta)$, it may be more
sensible to use a scheme only tempering the likelihood part:
\begin{equation}\label{eqn:likelihoodTempering}
  f_{(T)}(\theta) \;\propto\; \pi(\theta)\,\likeli(\theta)^\frac{1}{T},
\end{equation}
in which case $f_{(T)}(\theta)$ goes towards the prior~$\pi(\theta)$
for $T\rightarrow\infty$ \cite{Roever2007-thesis}.  Both uniform
distribution and prior distribution may in general not be the most
sensible choice, as was pointed out above, since the tempering is also
supposed to enhance the algorithm's \textsl{stochastic search}
properties.  Assume that one had a distribution~$\pstar(\theta)$
available, which leads to an optimal sampling (w.r.t. to some
pre-specified criterion), and which is then the desired density for
$T\rightarrow\infty$.  This suggests a generalized tempering
parametrisation:
\begin{equation}\label{eqn:parallelTemperingSetup}
  f_{(T)}(\theta) \;\propto\; \pstar(\theta) \left(\frac{f(\theta)}{\pstar(\theta)}\right)^\frac{1}{T}
                  \; = \; \pstar(\theta)^{1-\frac{1}{T}}\, f(\theta)^\frac{1}{T}
\end{equation}
which in the special cases of $\pstar(\theta)\propto 1$ and
$\pstar(\theta)=\pi(\theta)$ again yields the tempering schemes from
(\ref{eqn:plainTempering}) and (\ref{eqn:likelihoodTempering}) above.
The question now is how to choose such a limiting
distribution~$\pstar(\theta)$ based on given prior information and
template metric.

\section{\label{sec:DecisionTheory}The decision theoretic approach}
Let $g(\theta)$ be the determinant of the template metric as a
function of the signal parameters.  A large value of $g$ means that
that templates need to be densely spaced around~$\theta$, while a
smaller $g$ indicates that a coarser spacing is sufficient.  The
volume ``covered'' by a template placed at parameter~$\theta$ is
proportional to $g(\theta)^{-\frac{1}{2}}$, and hence the probability
density to sample from for setting up a random template bank is given
by $\rhohat(\theta)\propto\sqrt{g(\theta)}$
\cite{MessengerPrixPapa2009}. 

Now consider the case of the true parameter value
being~$\theta_0\in\Theta$.  The actual value~$\theta_0$ is unknown,
what is known is the prior probability density~$\pi(\theta)$.
Whenever a template~$\theta^\ast$ is placed in parameter space, it is
considered a \textsl{match} if it was sufficiently close to the true
value~$\theta_0$.  What exactly is ``sufficiently close'' is
determined via mismatch considerations and is expressed through the
template metric. Then the probability of a match is
\begin{equation}\label{eqn:successprob1}
  \prob\bigl(\mbox{match} \,|\,\theta^\ast \bigr)
  \; = \;
  c \, \frac{1}{\sqrt{g(\theta^\ast)}} \, \pi(\theta^\ast),
\end{equation}
where $c\in\realline^+$ is a constant depending on how close a match
actually is required to be.  If one was to pick a \textsl{single}
template~$\theta^\ast$, the chances for success would obviously be
maximal where the above product reaches its maximum.  Analogously,
consider the case of a \textsl{given} true value~$\theta_0$ and
repeated, independent ``guesses'' drawn from $\pstar(\theta)$.  Then
for each single guess the probability of success is
\begin{equation}\label{eqn:successprob2}
  \prob\bigl(\mbox{match} \,|\,\theta_0 \bigr)
  \; = \;
  c \, \frac{1}{\sqrt{g(\theta_0)}} \, \pstar(\theta_0).
\end{equation}

What is desired is a distribution $\pstar$ from which to generate
independent draws so that the chances of getting a match are
``optimal''.  Whether or when one will get a match is a matter of
chance, depending on both the true value~$\theta_0\in\Theta$ and the
choice of~$\pstar\in\pstarspace$, where $\pstarspace$ is the space of
probability distributions over~$\Theta$.  Suppose we are interested in
minimizing the expected number of trials~$T$ (or \textsl{waiting
time}) until the first match.  Any choice of $\pstar$ implies a
probability distribution for~$T$; for a \textsl{given} true
value~$\theta_0$ and a sampling distribution~$\pstar$, $T$~follows a
geometric distribution with density and expectation:
\begin{equation}\label{eqn:geometricDensity}
  \prob\bigl(T\!=\!t \,|\,\theta_0 \bigr)
  \; = \;
  \Bigl(1\!-\!c \, \frac{1}{\sqrt{g(\theta_0)}} \, \pstar(\theta_0)\Bigr)^{t-1}\;
  \Bigl(c \, \frac{1}{\sqrt{g(\theta_0)}} \, \pstar(\theta_0)\Bigr),
\quad
  \expect\bigl[T \,|\,\theta_0\bigr]
  \; = \;
  \frac{1}{c \, \frac{1}{\sqrt{g(\theta_0)}} \, \pstar(\theta_0)}.
\end{equation}

In decision theoretic terms, we are given a \textsl{state-of-nature
space}~$\Theta$, an \textsl{action space}~$\pstarspace$, and a
\textsl{loss function}~$L:\Theta\times\pstarspace\rightarrow\realline$
with $L(\theta_0,\pstar)=\expect_{\pstar}[T|\theta_0]$
\cite{Berger,Ferguson}.  An optimal choice of $\pstar$ may now be
determined by minimizing the expected loss; integrating over the
possible values that $\theta_0$ could take, that (prior) expectation
is
\begin{equation}\label{eqn:priorExpectation}
  \expect[T]
  \; = \;
  \frac{1}{c} \int_\Theta \frac{1}{\frac{1}{\sqrt{g(\theta_0)}} \, \pstar(\theta_0)} \,\pi(\theta)\,\differential \theta,
\end{equation}
which is minimized by choosing
\begin{equation}\label{eqn:optimalPstar}
  \pstar(\theta)
  \; \propto \;
  \sqrt{\pi(\theta)\,\sqrt{g(\theta)}}
  \; = \;
  \sqrt{\pi(\theta)\,\rhohat(\theta)},
\end{equation}
i.e., the optimal~$\pstar$ here is proportional to the geometric mean
of $\pi$ and $\rhohat$, and independent of~$c$.

The distribution defined through the density~$\rhohat$ that is usually
utilized for random template banks \cite{MessengerPrixPapa2009} plays
a particular role in this context.  From equation
(\ref{eqn:successprob2}) one can see that by setting $\pstar:=\rhohat$
the probability of a match (and with that also the waiting time)
becomes independent of the actual parameter value~$\theta_0$, so that
$\rhohat$ constitutes an \textsl{equalizer rule}.  From
(\ref{eqn:optimalPstar}) it follows that $\rhohat$ will be optimal in
the case that the prior happens to be $\pi\!=\!\rhohat$.  This implies
that $\pi\!=\!\rhohat$ defines the ``\textsl{least favourable prior
distribution}'' for this case, and that $\pstar\!=\!\rhohat$ also
constitutes the \textsl{minimax} strategy (independent from the
particular prior~$\pi$), as it minimizes the maximum of
$\expect\bigl[T \,|\,\theta_0\bigr]$ across all possible true
values~$\theta_0$ \cite{Berger}.  Since $\pstar\!=\!\rhohat$ leads to
a uniform match probability in~(\ref{eqn:successprob2}), it actually
constitutes the equalizer rule for the wider family of optimality
criteria that are functions of $\prob(\mbox{match} \,|\,\theta_0 )$.

\section{\label{sec:examples}Examples}
\subsection{Toy example 1: Gaussian prior}\label{sec:toyExample1}
Consider a parameter space $\Theta=\realline$ where the prior is
Gaussian with mean~$\mu$ and variance~$\sigma^2$:
$\pi=\normaldistn(\mu,\sigma^2)$, and the template metric is
\textsl{flat}, i.e., $g(\theta)=\gamma$ is independent of $\theta$.
Then the equalizer rule~$\rhohat$ does not exist, and the optimal rule
would be $\pstar=\normaldistn(\mu, 2\sigma^2)$.

\subsection{Toy example 2: Numerical simulation}\label{sec:toyExample2}
\begin{figure}
\begin{center}
\includegraphics[width=0.98\textwidth]{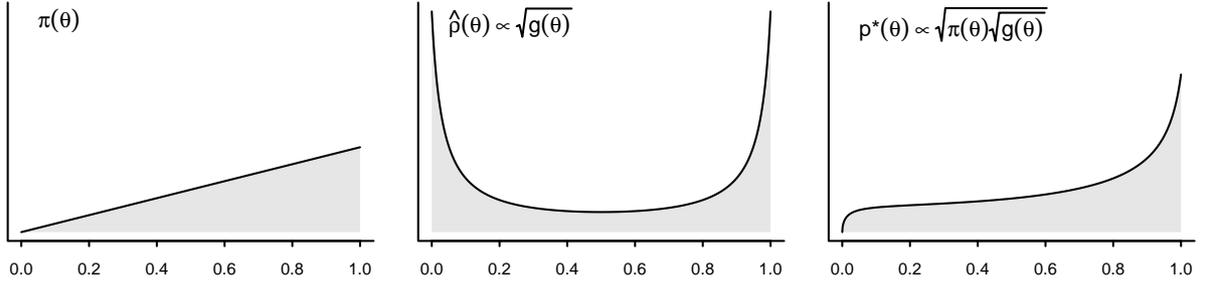}
\end{center}
\caption{\label{fig:toyExample1}Densities of the distributions $\pi$,
$\rhohat$ and $\pstar$ for the toy example discussed in
Sec.~\ref{sec:toyExample2}.}
\end{figure}

Consider a parameter space $\Theta=[0,1]$, where the prior and
template metric behave as shown in Fig.~\ref{fig:toyExample1}.
For this simple case the behaviour of different sampling strategies
can be simulated numerically, by drawing ``true'' parameter
values~$\theta_0$ from the prior distribution and then drawing
``guesses''~$\theta^\ast$ from either $\rhohat$ or $\pstar$ in order
to see how the strategies differ.

\begin{figure}[h]
\includegraphics[width=22pc]{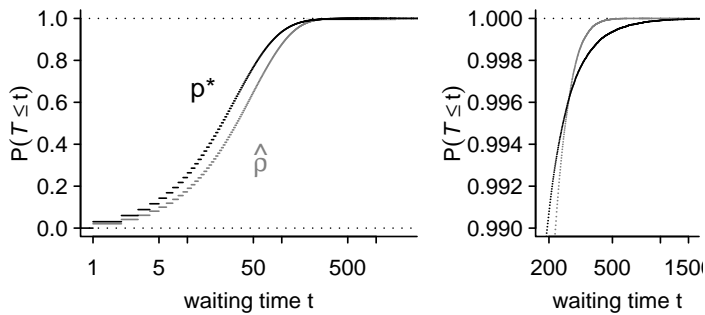}\hspace{2pc}%
\begin{minipage}[b]{12pc}\caption{\label{fig:toyExample2}Cumulative 
distributions of the resulting waiting times~$T$ when using 
sampling strategies~$\pstar$ and $\rhohat$ in the toy example of 
Sec.~\ref{sec:toyExample2}. The right panel shows a zoom-in on the 
differing tail behaviour.}
\end{minipage}
\end{figure}
Fig.~\ref{fig:toyExample2} illustrates the distribution of the
resulting times~$T$, for both the minimax and optimal strategies
$\rhohat$ and $\pstar$.  As expected, the average waiting time is
lower for $\pstar$, and one can see that the minimax strategy performs
better in the unlikely ``worst cases''.

\subsection{Binary inspiral example}
\begin{figure}
\begin{center}
\includegraphics[width=0.98\textwidth]{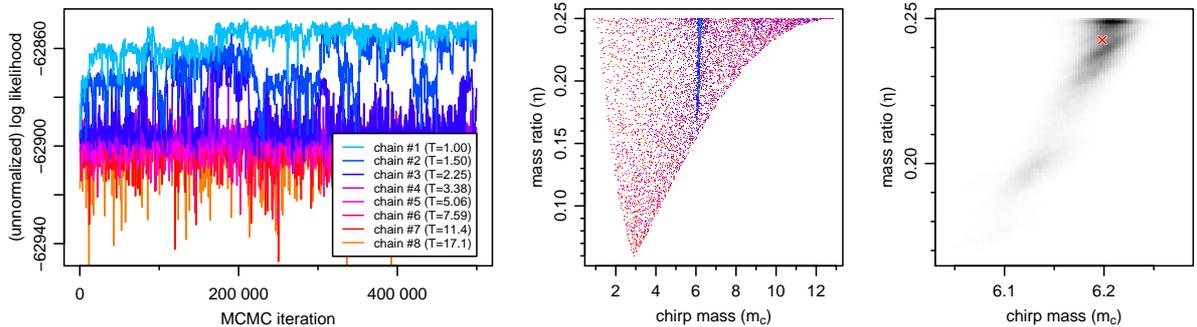}
\end{center}
\caption{\label{fig:PTPlots}This plot illustrates the behaviour of a
Parallel Tempering algorithm utilizing the distribution~$\pstar$ when
running on simulated data. The left panel shows how the algorithm's
`cool' chains manage to ascend to greater likelihood values while the
tempered chains keep sampling at lower likelihood values. The 2nd
panel is a scatter plot of mass parameter samples from all the
different chains (after the algorithm's burn-in phase). The right
panel eventually shows the resulting mass parameters' marginal
posterior density derived from the `cool' chain~\#1 alone; the cross
indicates the true parameter value.}
\end{figure}

The prior~$\pi$ and minimax sampling rule~$\rhohat$ for the mass
parameters of a binary inspiral event were shown in
Fig.~\ref{fig:inspiralParam}.  The right panel of the same figure also
shows the resulting optimized sampling distribution~$\pstar$. The
obvious discrepancy between least favourable~($\rhohat$) and actual
prior~($\pi$) suggests that there actually is a gain in doing the
optimization.  Fig.~\ref{fig:PTPlots} shows how a parallel tempering
algorithm for parameter estimation behaves when utilizing the
distribution~$\pstar$ for high-temperature chains as described in
Sec.~\ref{sec:ParallelTempering} (\ref{eqn:parallelTemperingSetup}).
The MCMC chains quickly converge to the true parameter values, while
the higher-temperature chains keep scanning the parameter space
efficiently.

\section{\label{sec:ConlcusionsOutlook}Conclusions and outlook}
We have applied a decision-theoretic approach in order to derive an
optimized sampling distribution to be used within a parallel tempering
MCMC implementation.  The optimization step here provides a natural
link between the parameter space metric and the prior information
about the parameter values.  The particular optimality criterion
chosen here (the expected time until a matching template is found,
$\expect[T|\theta_0]$) turns out to be computationally convenient, as
the resulting sampling distribution~$\pstar$ is independent of the
particular mismatch threshold~$c$, and is almost trivial to implement
within an MCMC application.  Other criteria are conceivable though,
like the probability of a missed detection within $N$ samples
$\prob(T\!>\!N|\theta_0)$ for example, which may then lead to more
complicated results.

The general approach used here should also be useful in other
contexts; it turns out that the distribution usually used for setting
up random template banks here constitutes the special case of a
\textsl{minimax} strategy, which implies that the explicit
specification of particular figures-of-merit and the consideration of
prior information may yield great efficiency improvements, especially
in cases where the implicitly assumed \textsl{least favourable prior}
greatly deviates from the actual prior information as in the binary
inspiral case.  In the framework discussed above, the resulting
optimized sampling distribution~$\pstar$ even exists for cases where
the minimax rule does not (as in the example of
Sec.~\ref{sec:toyExample1} above).  This suggests that a similar
approach may also make other ad-hoc fixes like the mass parameter
bounds in the binary inspiral example dispensable, as it would
naturally focus in on the promising parameter range while ruling out
too unlikely and too costly regions of parameter space.

\ack 
The author would like to thank Chris Messenger, Reinhard Prix and
Graham Woan for helpful discussions.
This work was supported by the Max-Planck-Society.

\section*{References}
\bibliographystyle{iopart-num}
\bibliography{/home/christian/literature/literature}

\end{document}